\documentclass[twocolumn]{aastex631}

\usepackage{float}
\usepackage{rotating}
\usepackage{mathtools}
\usepackage{mathastext}
\usepackage{changepage}
\usepackage{textcomp}
\usepackage{pgfplots}
\usetikzlibrary{plotmarks}
\pgfplotsset{compat=1.9}

\def\ps{\texttt{+}}
\def\ms{\texttt{-}}

\DeclarePairedDelimiter\abs{\lvert}{\rvert}

\shorttitle{NSC DR2 Comoving Systems}
\shortauthors{Kiwy et al. 2022}

\newenvironment{rotatepage}
    {\clearpage\pagebreak[4]\global\pdfpageattr\expandafter{\the\pdfpageattr/Rotate 90}}
    {\clearpage\pagebreak[4]\global\pdfpageattr\expandafter{\the\pdfpageattr/Rotate 0}}

\begin{document}

\title{Discovery of 34 low-mass comoving systems using NOIRLab Source Catalog DR2}

\author[0000-0001-8662-1622]{Frank Kiwy}
\affiliation{Backyard Worlds: Planet 9}
\author[0000-0001-6251-0573]{Jacqueline K. Faherty}
\affiliation{Department of Astrophysics, American Museum of Natural History, Central Park West at 79th Street, NY 10024, USA}
\author[0000-0002-1125-7384]{Aaron Meisner}
\affiliation{NSF's National Optical-Infrared Astronomy Research Laboratory, 950 N. Cherry Ave., Tucson, AZ 85719, USA}
\author[0000-0002-6294-5937]{Adam C. Schneider}
\affiliation{United States Naval Observatory, Flagstaff Station, 10391 West Naval Observatory Rd., Flagstaff, AZ 86005, USA}
\affiliation{Department of Physics and Astronomy, George Mason University, MS3F3, 4400 University Drive, Fairfax, VA 22030, USA}
\author[0000-0003-4269-260X]{J. Davy Kirkpatrick}
\affiliation{IPAC, Mail Code 100-22, Caltech, 1200 E. California Blvd., Pasadena, CA 91125, USA}
\author[0000-0002-2387-5489]{Marc J. Kuchner}
\affiliation{NASA Goddard Space Flight Center, Exoplanets and Stellar Astrophysics Laboratory, Code 667, Greenbelt, MD 20771, USA}
\author[0000-0002-6523-9536]{Adam J. Burgasser}
\affiliation{Center for Astrophysics and Space Science, University of California, San Diego, La Jolla, CA 92093, USA}
\author[0000-0003-2478-0120]{Sarah Casewell}
\affiliation{School of Physics and Astronomy, University of Leicester, University Road, Leicester, LE1 7RH, UK}
\author[0000-0003-2102-3159]{Rocio Kiman}
\affiliation{Kavli Institute for Theoretical Physics, University of California, Santa Barbara, CA 93106, USA}
\author[0000-0002-2682-0790]{Emily Calamari}
\affiliation{Department of Physics, Barnard College, Columbia University, New York, NY 10027, USA}
\author[0000-0003-2094-9128]{Christian Aganze}
\affiliation{Department of Physics, University of California, San Diego, La Jolla, CA 92093}
\author[0000-0002-5370-7494]{Chih-Chun Hsu}
\affiliation{Department of Physics, University of California, San Diego, La Jolla, CA 92093}
\author[0000-0003-4864-5484]{Arttu Sainio}
\affiliation{Backyard Worlds: Planet 9}
\author{Vinod Thakur}
\affiliation{Backyard Worlds: Planet 9}
\author{The Backyard Worlds: Planet 9 Collaboration}

\begin{abstract}
We present the discovery of 34 comoving systems containing an ultra-cool dwarf found by means of the NOIRLab Source Catalog (NSC) DR2. NSC's angular resolution of $\sim1\arcsec$ allows for the detection of small separation binaries with significant proper motions. We used the catalog's accurate proper motion measurements to identify the companions by cross-matching a previously compiled list of brown dwarf candidates with NSC DR2. The comoving pairs consist of either a very low-mass star and an ultra-cool companion, or a white dwarf and an ultra-cool companion. The estimated spectral types of the primaries are in the K and M dwarf regimes, those of the secondaries in the M, L and T dwarf regimes. We calculated angular separations between $\sim2$ and $\sim56\arcsec$, parallactic distances between $\sim43$ and $\sim261$ pc and projected physical separations between $\sim169$ and $\sim8487$ AU. The lowest measured total proper motion is 97 mas yr$^{-1}$, the highest 314 mas yr$^{-1}$. Tangential velocities range from $\sim23$ to $\sim187$ km s$^{-1}$. We also determined comoving probabilities, estimated mass ratios and calculated binding energies for each system. We found no indication of possible binarity for any component of the 34 systems in the published literature. The discovered systems can contribute to the further study of the formation and evolution of low-mass systems as well as to the characterization of cool substellar objects.
\end{abstract}


\section{Introduction}

Binary and multiple stellar systems have long been used to study the formation and fundamental properties of stars. With the first discovery of brown dwarf binary systems \citep{1998ApJ...509L.113M, 1999AJ....118.2460B, 1999Sci...283.1718M}, these investigations have been extended into the substellar regime. Physical properties of individual brown dwarfs (e.g. mass, metallicity or surface gravity) are  difficult to determine due to the absence of main sequence hydrogen burning, causing them to cool and fade over time. For brown dwarfs in multiple systems, however, physical properties, age and composition can be inferred from their stellar companions. Further, the binary fraction, mass ratio distribution and separation of brown dwarf companion systems can provide constraints on star formation and dynamical evolution \citep{2007AA...466..943G}.

One valuable type of benchmark systems are those with a wide binary composed of resolved companions, the primary being a main sequence star and the secondary being an L or T dwarf. Because it is very difficult to determine properties such as metallicity and age of L and T dwarfs, these properties can be inferred from the primary since comoving systems are assumed to have formed at the same time, from the same material and developed in the same environment. However, benchmark systems involving L or T dwarf companions are more difficult to find than systems composed of earlier type companions, because often either the primary is saturated or the secondary remains undetected. Moreover, the binary fraction seems to decrease from early to late primary spectral types \citep{2012ApJ...757..141K}. While the binary fraction for solar type stars ranges within 50-60\% \citep{1991AA...248..485D, 2010ApJS..190....1R}, it decreases to 30-40\% for M stars \citep{1992ApJ...396..178F, 2004ASPC..318..166D, 2014ApJ...789..102J, Winters2019}. For field brown dwarfs, the resolved binary fraction for very low-mass systems is around 10-20\% \citep{2003ApJ...587..407C, 2006ApJS..166..585B, 2011AJ....142...57G, 2015AA...578A...1H, 2018MNRAS.479.2702F}.

Identifying comoving companions has become easier through the use of large, multi-epoch surveys such as the Wide-field Infrared Survey Explorer \citep{2010AJ....140.1868W}, the {\it Gaia} Mission \citep{2016AA...595A...1G}, or the DESI Legacy Imaging Surveys \citep{2019AJ....157..168D}. Resulting catalogs like CatWISE2020 \citep{2021ApJS..253....8M}, {\it Gaia} DR2 \citep{2018AA...616A...1G}, {\it Gaia} EDR3 \citep{2021AA...649A...1G} or NSC DR2 \citep{2021AJ....161..192N} are excellent resources for finding late-type moving objects and their potential companions by using the provided proper motions. Two objects sufficiently close to each other on the sky and having comparable proper motions can provide a basis for inferring a common origin.

This paper is outlined as follows. In Section \ref{sec:nsc}, we briefly describe the NSC and its key elements relevant to this work. In Section \ref{sec:method}, we present our search method, and in Section \ref{sec:systems}, we characterize the discovered systems. We discuss some of the systems in more detail in Section \ref{sec:discussion}. 

\section{NOIRLab Source Catalog DR2}\label{sec:nsc}

NSC DR2 \citep{2021AJ....161..192N} is based on public image data from the NOIRLab Astro Data Archive. These images come from telescopes in both hemispheres (CTIO$\ms$4m$\ps$DECam, KPNO$\ms$4m$\ps$Mosaic3 and Bok$\ms$2.3m$\ps$90Prime) and cover $\sim35,000$ square degrees of the sky. A significant part of the images were obtained by the Dark Energy Survey \citep{2018ApJS..239...18A} and the DESI Legacy Imaging Surveys \citep{2019AJ....157..168D}. NSC DR2 includes more than 3.9 billion single objects with over 68 billion individual source measurements. It has depths of $\sim23rd$ magnitude in most broadband filters (u, g, r, i, z, Y and VR), accurate proper motions and an astrometric accuracy of $\sim7$ mas.

NSC provides proper motion measurements that push much fainter at optical wavelengths than {\it Gaia} DR2 or EDR3. At g-band, NSC is $\sim2.5$ magnitudes deeper than {\it Gaia} and allows proper motion searches for distant stars with high tangential velocities over a volume $\sim25$ times larger than {\it Gaia}. NSC can also measure motions for white dwarfs much fainter than those detected by {\it Gaia}, expanding the census of white dwarfs in the solar neighborhood. NSC's accurate proper motions enable the discovery of ultra-cool white dwarf binaries where metallicity and radial velocity can be derived from a main sequence companion \citep{2020MNRAS.493.6001L}.

Due to its excellent red-optical sensitivity and sky coverage, NSC provides many new possibilities to search for ultra-cool stars and brown dwarfs in the solar neighborhood. CatWISE2020 \citep{2021ApJS..253....8M} represents one of the best infrared proper motion catalogs currently available. However, at its faint end, CatWISE motions are only significant above $\sim150-200$ mas yr$^{-1}$, unlike NSC, which can measure motions many times smaller at high significance. NSC's Y-band depth (23.4 mag) and angular resolution ($\sim1\arcsec$) allow for motion searches of late-type objects not possible with WISE. This enables queries for pairs of faint objects with consistent proper motions to find closely spaced low-mass companion systems.

\section{Search Method}\label{sec:method}

In a first step, we searched for previously missed L and T dwarfs in the NOIRLab Source Catalog using proper motion and photometry. We used the z, i and Y magnitudes along with the relations described in \cite{2019MNRAS.489.5301C} to determine the color cut for our initial selection. We required all objects having at least one detection in each of those three bands. The adopted color cut ($1.4 < (i - z)_{NSC} < 3.5$ and $0.4 < (z - Y)_{NSC} < 1.5$) should restrict the search to L and T dwarfs. However, it cannot be excluded that some late M dwarfs are within the selected objects. Note that both color constraints had to be satisfied for this selection. No quality cuts were applied to the photometry to avoid eliminating more distant late-type/faint objects with less accurate photometry. Candidates having proper motions below 100 mas yr$^{-1}$ have been discarded from the selection to provide proper motions significant enough to be visually confirmed by blinking images of different epochs. We avoided the galactic plane ($\abs{b} > 15$\degr) to further reduce the number of sources with spurious proper motions. We required a high Star/Galaxy classifier of at least 0.7 to ensure that the search only returns objects with a point-like morphology. The Star/Galaxy classifier is a NSC catalog column which provides information on an object's probability of being either a galaxy or a star, based on its morphology. Selected objects had to have a time baseline of at least six months between the first and last observation. By applying these criteria, we obtained a total of 2896 candidate ultracool dwarfs, mostly in the L and T regimes.

In a second step, we examined a $100\arcsec$ radius around all 2896 objects using the NSC by comparing the proper motion components of each object to the proper motion components of the objects found in the defined radius. We applied a proper motion matching tolerance of 20 mas yr$^{-1}$ to find potential companions either of earlier or later type. Note that we exclusively used NSC DR2 proper motions for this purpose. As in the first step, we excluded objects having a Star/Galaxy classifier below 0.7 and a time baseline of less than six months. To further reduce the number of false positives, each identified companion was cross-matched with {\it Gaia} EDR3 to discard objects with a parallax below 2 mas. This resulted in 46 pairs with similar proper motions.

Among those 46 potential companions, we found four false positives (no object visible in WISE or DECaLS imagery), three likely false positives with significant differences in their proper motion components, and thirteen known pairs, one of which (NSC J2322\ms6151) is referenced in \cite{2019MNRAS.485.4423S} and \cite{2019AA...627A.167C}, the other twelve find mention in the SUPERWIDE catalog \citep{2020ApJS..247...66H}. The known pairs, which are listed in Table \ref{tab:systems} (Rediscovered systems), are not further described in this paper.

We repeated both steps described above using a slightly different color cut than in the first selection ($i_{NSC} > 99$ and $0.4 < (z - Y)_{NSC} < 1.5$). The goal of this second selection was to target specifically substellar sources and not more distant warmer stars. Since late-type brown dwarfs emit more radiation in the near infrared, an i-band dropout is a strong signature for detecting a cold compact source. We therefore dropped the $i - z$ color but used the same constraints for the $z - Y$ color than in the first selection. We also required selected objects explicitly to have no i-band photometry by adding a corresponding constraint ($i > 99$) to exclude any objects found in the first selection. After applying the first step of our search method, we obtained a total of 2520 objects. Step two resulted in three additional systems, \#2, \#21 and \#24, having secondaries with estimated spectral types of T5, T3 and T0, respectively.

Each object has been visually inspected to determine whether its proper motion components are consistent with those of its comoving companion by blinking images of different epochs. For this, we used \texttt{AstroToolBox} \citep{2022ascl.soft01002K}, which is a Java tool set with a graphical user interface allowing us to blink either unWISE coadds \citep{2017AJ....154..161M, 2018AJ....156...69M} of epochs 2010 and 2014-2020, or DECaLS cutouts \citep{2019AJ....157..168D} from DR5, DR7, DR8 and DR9. DECaLS cutouts have a resolution $10\times$ higher than unWISE coadds (0.27\arcsec/pixel vs. 2.75\arcsec/pixel), making it possible to identify the components of binary systems down to an angular separation of $\sim1\arcsec$. All objects have been carefully checked against at least one background star, showing no visible motion, to eliminate false positives produced by misaligned images.

To assess the effect of the photometric uncertainties on the selection, we calculated the mean photometric error associated with the $(i - z)_{NSC}$ and $(z - Y)_{NSC}$ colors of the secondaries in our sample, which is 0.035 mag for both of these colors (Table \ref{tab:phot_errors}). The used color cut ($1.4 < i - z < 3.5$ and $0.4 < z - Y < 1.5$) was determined by the means of the \cite{2019MNRAS.489.5301C} relations with the following color values for the given spectral types: \{L0, T9\} = \{($i - z = 1.48, z - Y = 0.43$), ($i - z = 3.39, z - Y = 1.5$)\}. Note that the Carnero Rosell relations do not give uncertainties on their color values. When we subtract the mean photometric errors from the color values for spectral type L0, we obtain the following values: $i - z = 1.48 - 0.035 = 1.445$ and $z - Y = 0.43 - 0.035 = 0.395$. While the value for the $i - z$ color is still above the lower limit of our color cut, that for the $z - Y$ color is 0.005 mag below the limit. Given this small color value difference, we conclude that it has minimal, if any, effect on the selection. As for the upper limits of the used color cut, the photometric errors do not affect the selection since our sample does not contain objects with estimated spectral types later than T5. This corresponds to the color values $i - z = 3.25$ and $z - Y = 1.02$ from the Carnero Rosell relations, which both are well below the upper bounds of the employed color cut ($i - z = 3.5, z - Y = 1.5$), and stay well below even if we add the mean photometric uncertainties  ($i - z = 3.25 + 0.035 = 3.285$ and $z - Y = 1.02 + 0.035 = 1.055$).

Aside from the 29 new systems discovered through the method described above, an additional five new systems (\#3, \#11, \#16, \#25 and \#27) were recovered serendipitously when visually checking a list of objects from a previous NSC DR2 search focused on high proper motion ultra-cool dwarf candidates. The color and proper motion cuts used in that search were somewhat different from those applied in step one of this work. In each of these cases, the companion could be identified through \texttt{AstroToolBox}'s image blinker. For those five systems, either the primary or the secondary did not satisfy all the constraints defined in the search method of this work.

At the same time these searches were on-going, several members of the Backyard Worlds: Planet 9 citizen science project \citep{Kuchner2017} were searching for similar cold compact objects via the Zooniverse portal or via a research scientist guided side project. Arttu Sainio recovered system \#21, Vinod Thakur recovered systems \#13, \#20, and Sam Goodman recovered systems \#1 and \#35. Details on those searches and additional discoveries by the larger Backyard Worlds team will be reported in a forthcoming paper.

\begin{figure}[H]
\centering
\includegraphics[width=18cm]{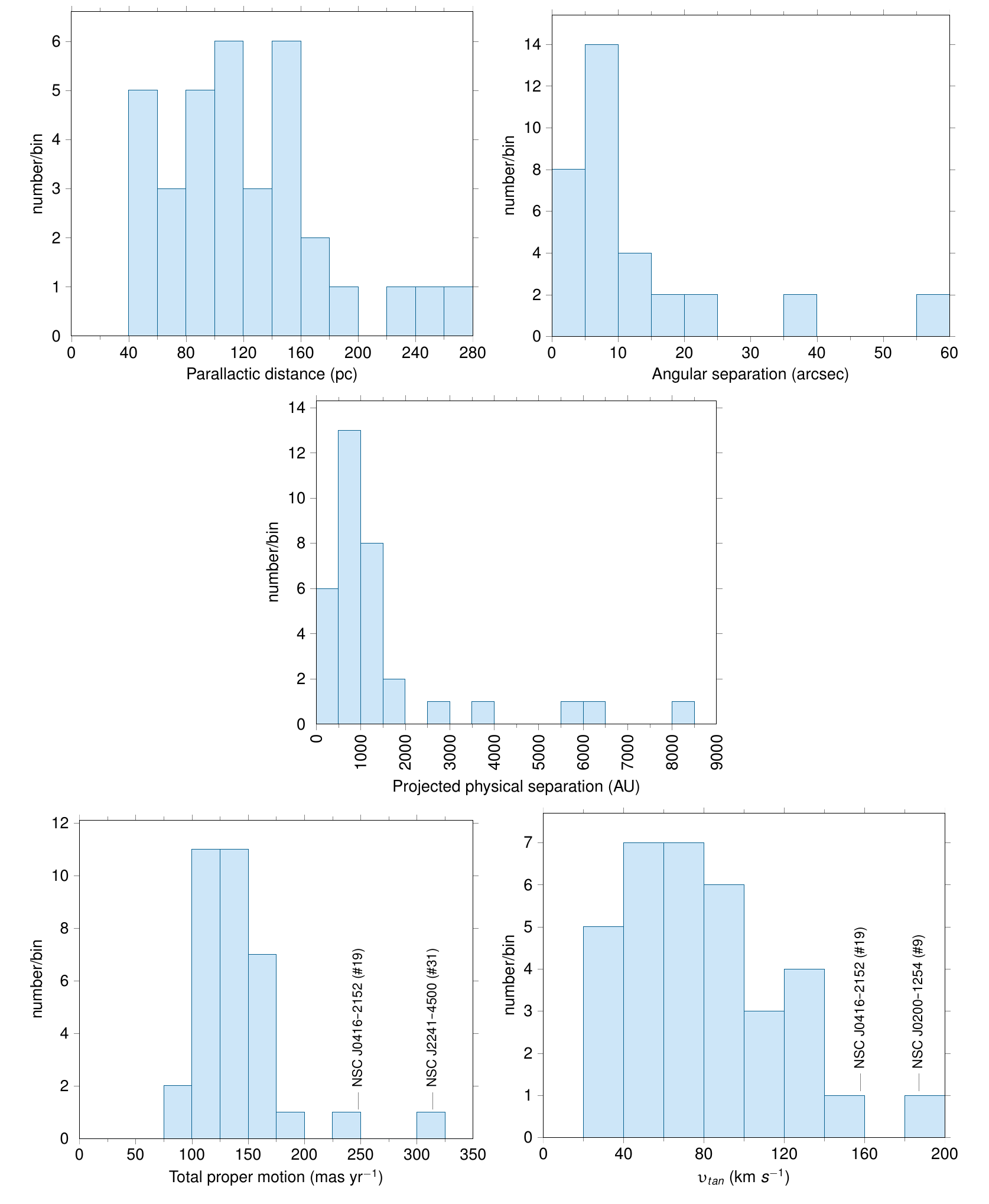}
\caption{Distributions of parallactic distance, angular separation, projected physical separation, total proper motion and $\upsilon_{tan}$ from Tables \ref{tab:systems} and \ref{tab:nsc_astro}.}\label{fig:histograms}
\end{figure}

We identified a total of 34 new candidate binary systems, most of which are located in the \href{https://datalab.noirlab.edu/des}{DES footprint} \citep{2021ApJS..255...20A}. The general properties of these systems are given in Table \ref{tab:systems} (Newly discovered systems) with additional astrometry from NSC DR2 in Table \ref{tab:nsc_astro} and relevant photometry in Table \ref{tab:photometry}. Histograms showing the distribution of distances, angular and physical separations, total proper motions and tangential velocities can be found in Figure \ref{fig:histograms}.

\section{Characterizing Systems}\label{sec:systems}

\subsection{Spectral Type Estimates}

We evaluated the photometric quality of our sample by calculating the weighted mean of all photometric errors in all bands of the employed photometric systems. Since the secondaries are generally fainter and often blended by the primaries, we calculated the weighted mean separately and found that it is 0.015 mag for the primaries and 0.078 mag for the secondaries (Table \ref{tab:phot_errors}). We deduce that the overall photometric quality of our sample is good and even excellent for most of the primaries but degrades for the secondaries (with regard to the primaries).

\begin{figure}[H]
\centering
\includegraphics[width=18cm]{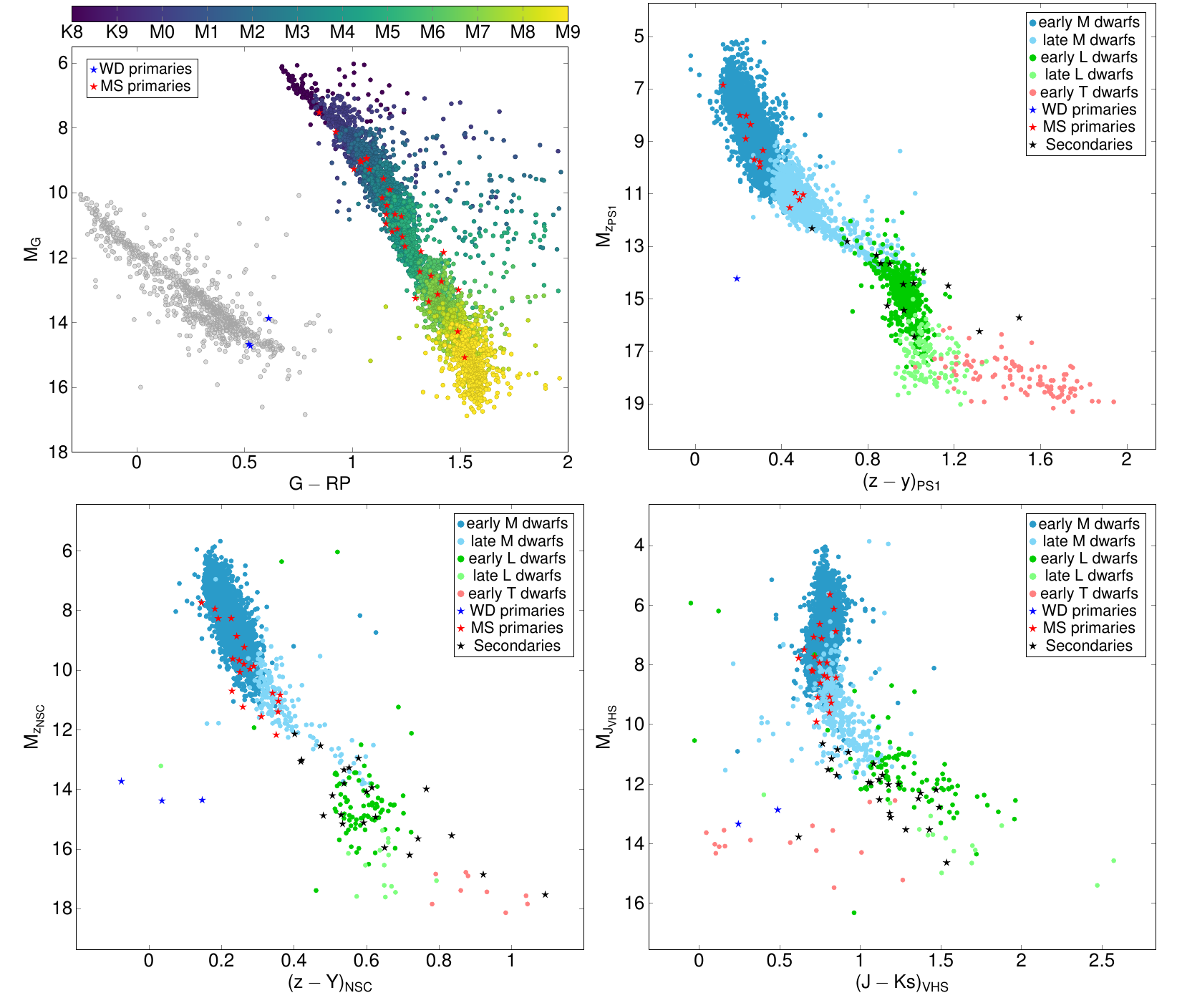}
\caption{Top left panel: {\it Gaia} CMD of the primaries. Top right panel: Pan-STARRS CMD; Bottom left panel: NSC CMD; Bottom right panel: VHS CMD. Five-pointed star symbols: the blue stars represent the white dwarf (WD) primaries, the red stars correspond to the main sequence (MS) primaries, and the black stars refer to the secondaries. We used the parallax of the primaries to calculate the absolute magnitude of the secondaries.}\label{fig:cmds}
\end{figure}

We plotted different types of color-magnitude diagrams (CMDs), using {\it Gaia} EDR3 G-band versus $G-G_{RP}$ color, Pan-STARRS DR2 z-band versus $z-y$ color, NSC DR2 z-band versus $z-Y$ color, and VISTA VHS DR5 J-band versus $J-Ks$ color (Figure \ref{fig:cmds}). For the {\it Gaia} CMD, we plotted the primaries on a comparative sample composed of late K and M dwarfs (K8-M9) and DA white dwarfs extracted from the SIMBAD database \citep{2000A&AS..143....9W}. For the Pan-STARRS, NSC and VISTA VHS CMDs, we used as a comparative sample the M dwarfs from \cite{2018ApJS..234....1B} and the L \& T dwarfs from \cite{UltracoolSheet}. The M dwarfs were cross-matched with {\it Gaia} EDR3 to obtain the parallaxes to calculate the absolute magnitudes. M, L and T dwarfs were cross-matched with NSC and VISTA VHS to get the corresponding photometry. A radius of $2\arcsec$ was used for both cross-matches. We split the comparative sample into early and late-type M, L and T dwarfs and used a different color for each of these categories. We always plotted all the systems having the corresponding photometry on each of those CMDs.

According to our color-magnitude diagrams, we found 31 systems composed of a main sequence star and an ultra-cool companion, and 3 systems composed of a white dwarf and an ultra-cool companion. The estimated spectral types of the main-sequence stars are in the late-K and M dwarf regimes, those of the ultra-cool companions in the late-M, L and early-T dwarf regimes. Our search resulted in 1 K$\ps$L, 11 M$\ps$M, 16 M$\ps$L, 3 M$\ps$T, 1 WD$\ps$M and 2 WD$\ps$L systems. More accurate spectral type estimates can be found in Table \ref{tab:systems}.

The spectral type estimates were obtained from \texttt{AstroToolBox}'s Photometric Classifier, which uses the available photometry along with the relations from \cite{2019AJ....157..231K}, \cite{2019MNRAS.489.5301C}, \cite{2018ApJS..234....1B} and \cite{Mamajek2021} to establish a spectral type classification by counting the occurrences of each determined spectral type. \cite{2019AJ....157..231K} relates to {\it Gaia} photometry for spectral types M0 to L7, \cite{2019MNRAS.489.5301C} relates to DES, VHS and AllWISE photometry for spectral types M1 to T9, and \cite{2018ApJS..234....1B} relates to Pan-STARRS, 2MASS and WISE photometry for spectral types M0 to T9. The \cite{Mamajek2021} relations are using {\it Gaia} photometry for spectral types B9-L8, Sloan photometry for spectral types M0-T8, 2MASS photometry for spectral types O9-Y1, and WISE photometry for spectral types B5-K5 and M5-Y4. \texttt{AstroToolBox} does not account for photometric uncertainties, but relies on the number of equal spectral type estimates from the various photometric systems and corresponding relations. For each object, we retained the spectral type with the most occurrences. In case there were two or more distinct spectral types with the same number of occurrences, we kept the one whose photometric distance best fits the {\it Gaia} parallax. When considering the mean photometric error of our sample in each of the used colors (Table \ref{tab:phot_errors}), we infer a spectral type uncertainty of about one sub-type for the primaries and two sub-types for the secondaries.

\subsection{Spectroscopy}

\subsubsection{Kast Optical Spectrograph}
Optical spectroscopy was obtained for the primary of system \#8 (NSC~J0153$-$0015A) using the Kast Double Spectrograph mounted on the Lick 3m Shane Telescope on 2021 Dec 12 (UT). Conditions were cloudy with average seeing of 1$\farcs$3. Two exposures of 1500~s each were obtained in the red channel using the 600/7500 grating and 1$\farcs$5-wide slit, providing 6000--9000~{\AA} at an average resolution of $\lambda/\Delta\lambda$ = 1900. Only one of the exposures was used due to the deteriorating weather conditions. The flux standard Feige 110 was observed at the start of the night for flux calibration, and the G2~V star HD 13043 was observed after NSC~J0153$-$0015A to measure telluric absorption. Data were reduced and analyzed using the \texttt{kastredux} package\footnote{\url{https://github.com/aburgasser/kastredux}} with default settings. The reduced spectrum, shown in Figure \ref{fig:spectra} (bottom left panel), has an average signal-to-noise of $\approx$25, due to the poor transparency. Nevertheless, we detect an overall red spectral energy distribution and weak molecular features consistent with an early-type M dwarf. We compared this spectrum to M dwarf optical template spectra from \cite{2007AJ....133..531B}, finding a best match to the M1 template, which we adopt for the classification of this primary.

\subsubsection{SpeX Infrared Prism Spectroscopy}
\begin{sidewaysfigure}
\centering
\includegraphics[width=\textwidth]{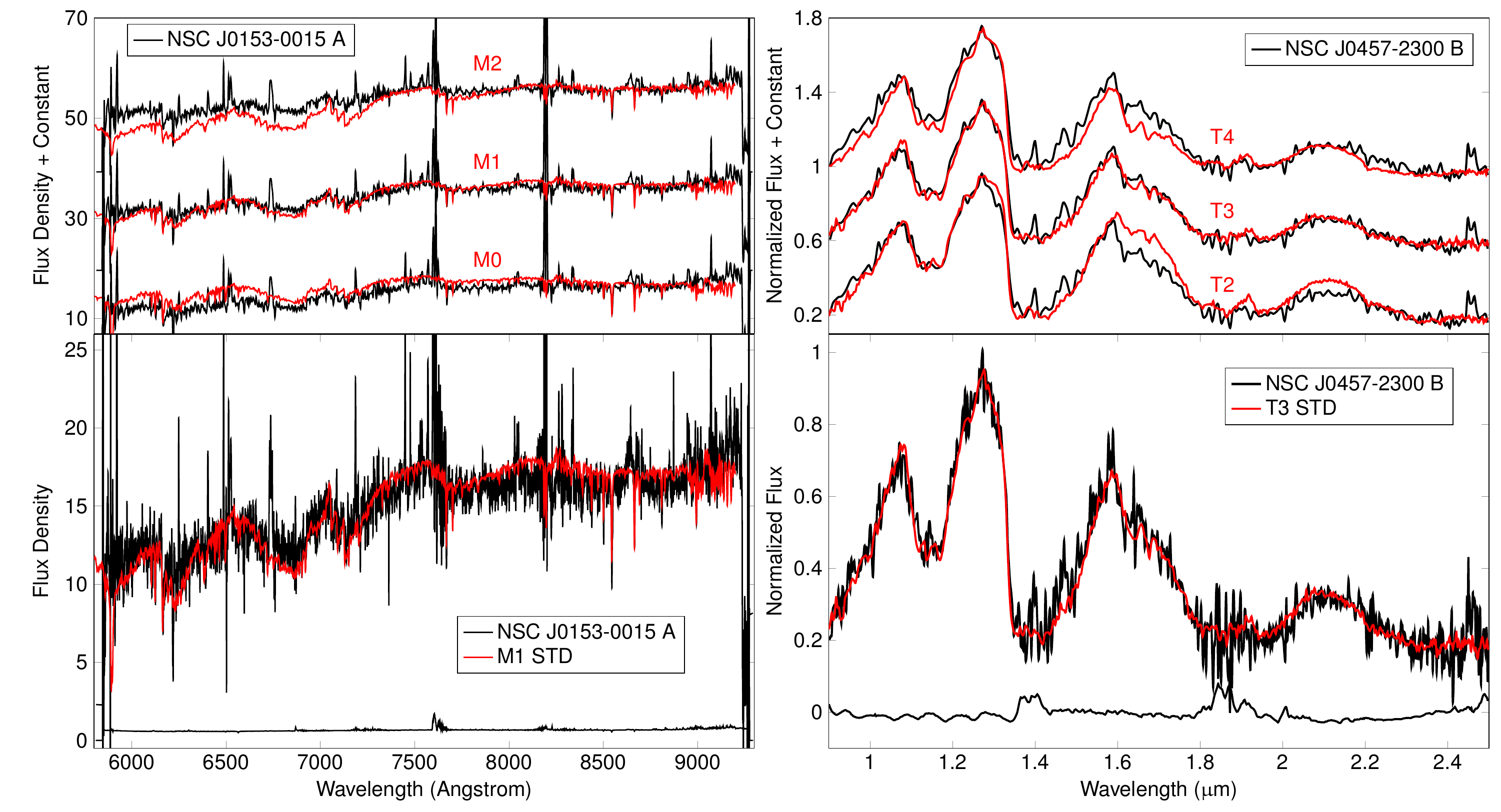}
\caption{Top left panel: Optical spectrum of NSC J015340.10-001550.34 A (primary of system \#8) compared to three M dwarf standards from \cite{2007AJ....133..531B}. The spectrum of our primary (smoothed for a better comparison) fits best with the M1 standard. Bottom left panel: Unsmoothed version of the same spectrum with the M1 standard overplotted. Top right panel: Near-infrared spectrum of NSC J045724.25-230012.74 B (secondary of system \#21) compared to three T dwarf standards from \cite{2004AJ....127.2856B}. The spectrum of our secondary (smoothed for a better comparison) fits best with the T3 standard. Bottom right panel: Unsmoothed version of the same spectrum with the T3 standard overplotted.}\label{fig:spectra}
\end{sidewaysfigure}

We observed the secondary in system \#21 using the SpeX spectrograph on NASA's IRTF telescope. The spectrum was taken on the night of 23 October 2021 (UT), under good conditions with minimal cloud coverage, in prism mode using the 0.8$\arcsec$ slit to achieve a resolving power of $\sim$100 - 500 over the 0.8 - 2.5~{$\mu$m} coverage. We obtained 7 AB nods using 180~s exposures on the target and then acquired the A0 star HD 32855 for telluric correction using 1~s exposures and 10 AB nods. All data were reduced using the \texttt{Spextool} package \citep{Cushing04} with telluric correction and flux calibration of the A0 stars following the technique described in \citet{Vacca03}. In Figure \ref{fig:spectra} (top right panel) we compare this spectrum to several T dwarf standards from \cite{2004AJ....127.2856B} and find a good agreement with the T3 standard, which confirms our spectral type estimate for this object.

\subsection{Distance Estimates}

\begin{figure}[H]
\includegraphics[width=8.5cm]{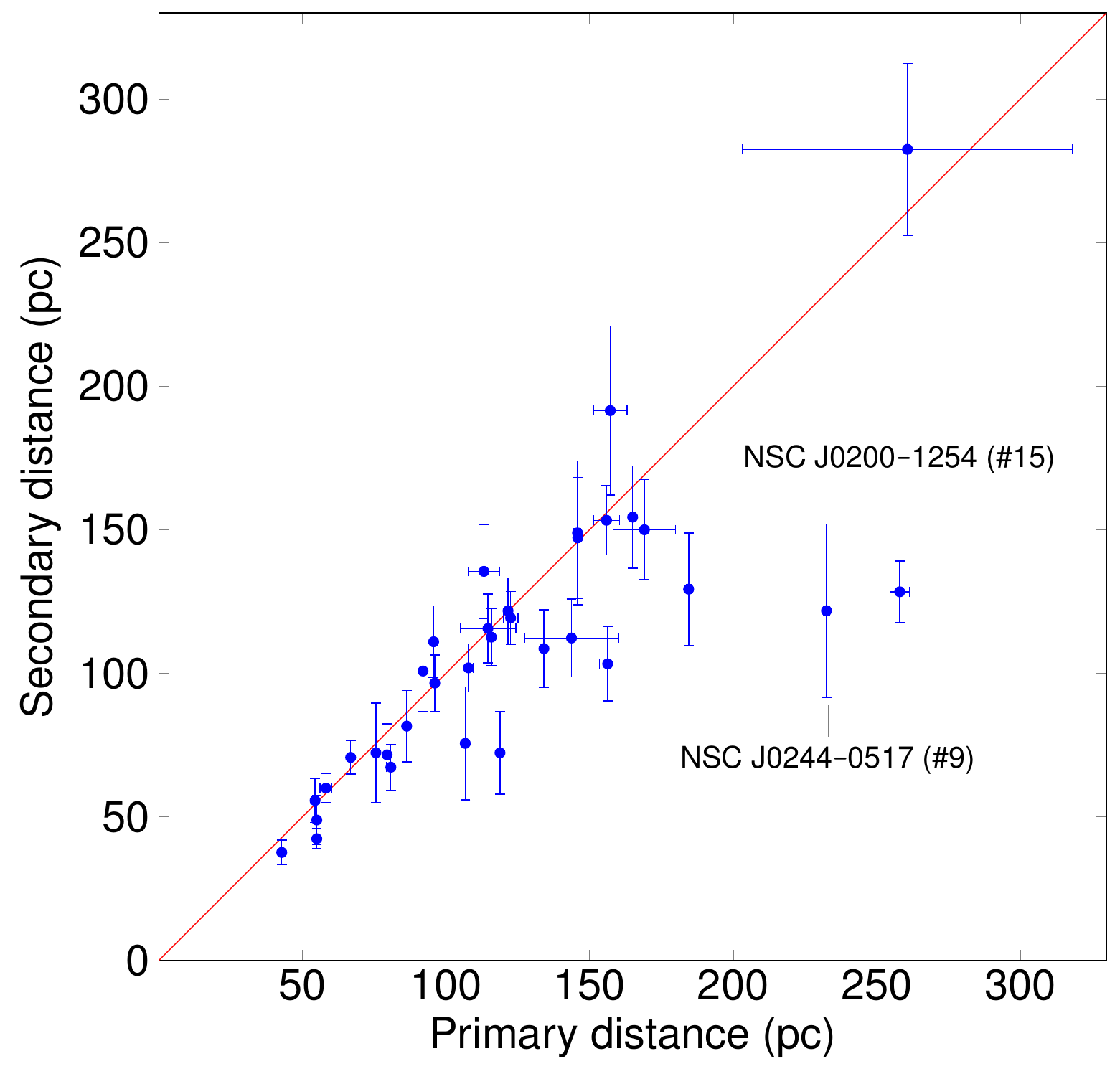}
\caption{Secondary mean photometric distance versus primary parallactic distance.} \label{fig:distances}
\end{figure}

Photometric distances were calculated for the secondaries by employing relevant spectro-photometric distance conversions. We used the spectral type estimates from \texttt{AstroToolBox} and derived the corresponding absolute magnitudes from Table 4 in \cite{2019AJ....157..231K} and Table 6 in \cite{2018ApJS..234....1B}. For each photometric system, we calculated distances using the bands covered by the above relations and retained the minimum and maximum values of the photometric system with the best match. 
For a given magnitude, we estimated an average distance difference between a given sub-type (e.g. L5) and its lower and upper neighbors (L4, L6) of $\sim20\%$ of the sub-type’s (L5) photometric distance. These estimations were done by comparing the distances of the lower and upper neighbors of a given sub-type at different magnitudes and in different bands. Since we assume a spectral type uncertainty of two sub-types for the secondaries, this can lead to distance discrepancies of up to $\sim40\%$ of the secondary’s photometric distance, depending on the quality of the photometry used for the spectral type estimates.
A comparison between the parallactic distance of the primaries and the mean photometric distance of the secondaries is presented in Table \ref{tab:comov_prob}, which also contains the bands used to calculate the minimum and maximum photometric distances as well as the spectral types employed to derive the absolute magnitudes for those calculations. We have illustrated the distance difference between the primaries and the secondaries in Figure \ref{fig:distances}. 
Note that for systems \#9 and \#15, the parallactic distance of the primary is about twice the photometric distance of the secondary, which we discuss further in Section \ref{sec:discussion}.

\subsection{Chance Alignment}\label{sec:chance_align}

We determined comoving probabilities using the CoMover code from \cite{2021ascl.soft06007G}, which uses the sky position, proper motion, parallax and optionally the heliocentric radial velocity of the host star with their respective errors, and compares these with the observables of the potential companion on the basis of various star models.

We executed the code once without a distance constraint and once with a distance constraint on the secondaries, since most photometric distances of our secondaries are based on spectral type estimates only. Also, the blending caused by some of the primaries can lead to inaccurate photometry, affecting spectral type and photometric distance estimates.

The resulting comoving probabilities are shown in Table \ref{tab:comov_prob}. Those without a distance constraint are all $>=$ 99.3\% except for system \#13, which only has a comoving probability of 73.6\%. Those with a distance constraint are $>=$ 99.8\% except for three systems (\#9, \#13 and \#15), which all have comoving probabilities significantly lower than 100\% (59.8, 87.0 and 4.5\% respectively).

We used the sky positions and proper motions with their respective errors from NSC DR2 for the primaries and secondaries. The parallax and associated error of the primaries are from {\it Gaia} EDR3. The parallax and corresponding error of the secondaries were inferred from their mean photometric distance, based on the spectral type estimates from \texttt{AstroToolBox}. As seen previously, a difference of two sub-types corresponds to a distance discrepancy of $\sim40\%$. To determine how much the comoving probabilities change with a difference of two sub-types, we increased the photometric distance of the secondaries by 40\% and re-ran the \texttt{CoMover} code. We found that most systems, which already had a high comoving probability, still have a high one after increasing the distance (see Table \ref{tab:comov_prob} columns 3 vs. 4).

\subsection{Literature Information}

We searched the literature for any derived physical properties for the primaries and secondaries of the newly discovered systems. All of these properties are from either \cite{2019AJ....158..138S} or \cite{2021MNRAS.508.3877G} and are given in Table \ref{tab:fund_param} along with all other literature references. Fundamental parameters were not available for any of the secondaries in our new systems. However, two of them find mention in \cite{2016AA...589A..49S}, which does not contain physical properties. Some of our primaries are referenced by one or more of the following literary sources: \cite{2018AA...619L...8R}, \cite{2003AA...397..575P}, \cite{1995yCat.1098....0L}, \cite{1976ApJS...30..351E}, \cite{1976ApJ...204..101E}, \cite{2003ApJ...582.1011S}, \cite{1996AAS..115..481W}, \cite{2016ApJS..224...36K}. None of these works include fundamental parameters.

\subsection{Astrometry}

We calculated distances and tangential velocities for all our systems using the {\it Gaia} EDR3 astrometry of the primaries (Table \ref{tab:systems}). Position angles, angular and physical separations were calculated by correcting the position of the primary for proper motion to the epoch of the secondary. NSC DR2 positions and mean observation epochs were used for this purpose. All separations were calculated using the general case equation $\theta =\cos ^{-1}\left[\sin \delta _{A}\sin \delta _{B}+\cos \delta _{A}\cos \delta _{B}\cos(\alpha _{A}-\alpha _{B})\right]$ to account for systems having larger RA coordinate separations at high declination. In Table \ref{tab:nsc_astro}, we computed the difference in proper motion between the primaries and secondaries, taking into account the errors in $\mu_{\alpha}$ and $\mu_{\delta}$. We also found that there is good agreement between the {\it Gaia} and NSC proper motions of the primaries (see {\it Gaia} proper motions in Table \ref{tab:systems} versus NSC proper motions in Table \ref{tab:nsc_astro}). Note that none of our secondaries has proper motions in {\it Gaia}.

\subsection{Mass Ratios and Binding Energies}

\begin{figure}[H]
\centering
\includegraphics[width=14cm]{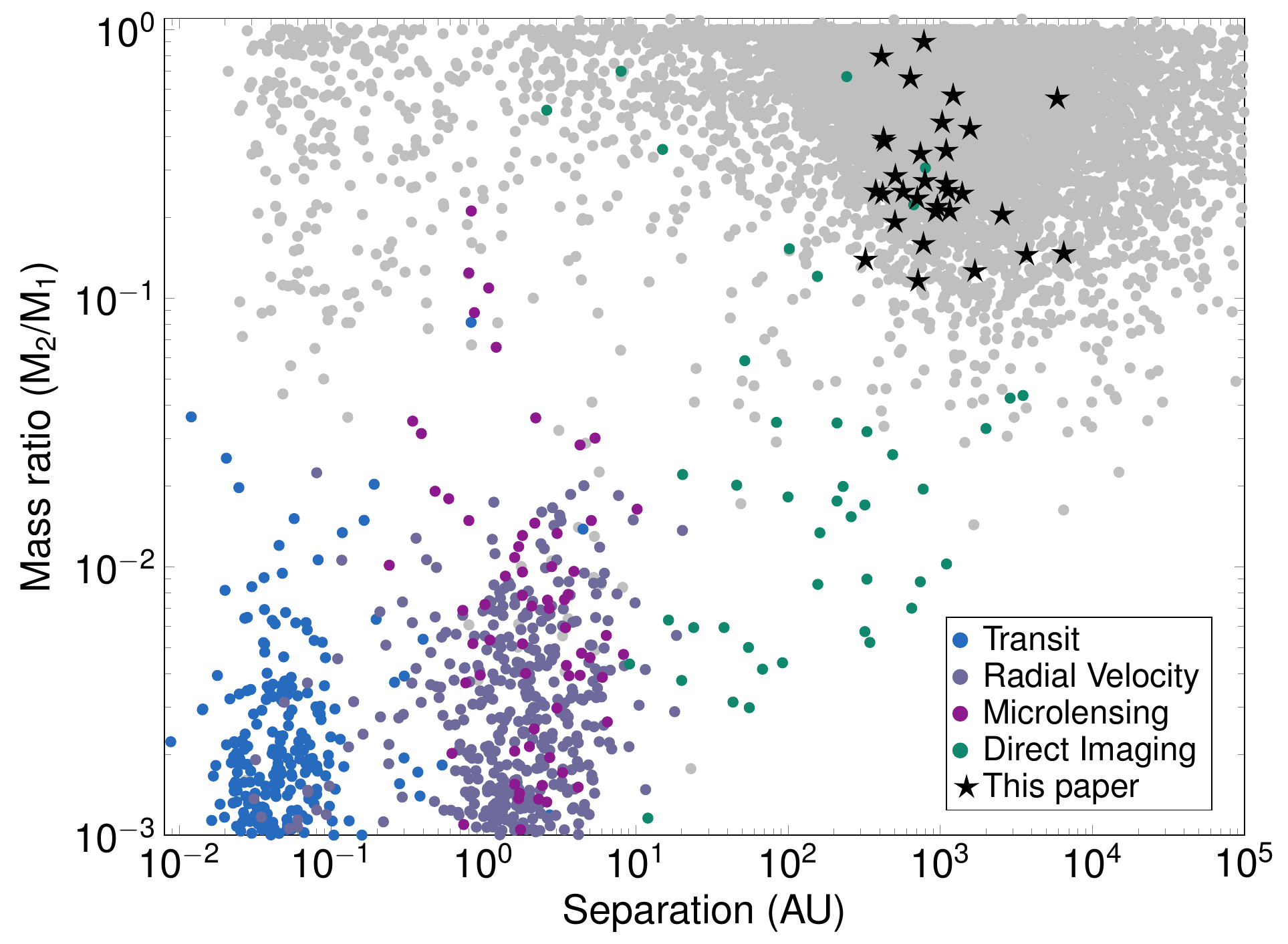}
\caption{Mass ratio (secondary/primary) versus projected separation for literature sources taken from \citealt{Faherty2020} and \citealt{2021ApJ...923...48F} (various color coded filled circles) and companions from this paper (black five-pointed stars). We have color coded the exoplanet detections (as defined by the Exoplanet Archive) by the detection method of the source.}\label{fig:mass_ratios}
\end{figure}

To compute total masses, mass ratios and binding energies for each systems (Table \ref{tab:mass_ratios}), we estimated component masses using a combination of literature references and relations described below. Table \ref{tab:fund_param} lists masses for the primary stars. For mass estimates of the secondaries we used either the \cite{Mamajek2021} relation for sources with absolute G magnitude $<$ 17, or we extrapolated from the spectral type to dynamical mass values computed in \cite{Dupuy2017}. In Figure \ref{fig:mass_ratios}, we overplotted our systems with the collection of very low-mass systems compiled in \cite{Faherty2020} as well as the collection of binaries from \cite{2021ApJ...923...48F}. We have omitted any system with a comoving probability $<$ 98\% (\#9, \#13 and \#15). The majority of our sources lie near the locus of field comoving sources as they have estimated total masses $>$ 0.2 M$_{\odot}$. Three systems (\#2, \#11 and \#14) have estimated total masses $<$ 0.2 M$_{\odot}$ as they contain low-mass secondaries (T5, L8, L4 respectively) with low-mass host stars ($\sim M7$). These three systems are at the low end of the binding energy distribution for field companion systems but still well contained within the locus of known sources.

\subsection{Age and Mass Estimates for the White Dwarfs}

We performed age, initial and final mass estimates for the three white dwarf primaries (systems \#25, \#27 and \#33) included in our sample. We used \texttt{wdwarfdate}\footnote{\url{https://github.com/rkiman/wdwarfdate}} (Kiman et al. in prep.), which is an open source code that estimates ages of white dwarfs in a Bayesian framework. It runs a chain of models to estimate the ages and their uncertainties from an effective temperature and a surface gravity. The code determines the age of the progenitor star, the cooling age of the white dwarf, and the total age. The age of the ultra-cool companion can be inferred directly from the total age of the white dwarf, provided that both formed at the same time. \texttt{wdwarfdate} also estimates the initial mass (that of the progenitor star) and the final mass (that of the white dwarf). The code uses the cooling tracks from \citep{2020ApJ...901...93B} to estimate the parameters of the white dwarfs, and MIST stellar evolutionary tracks computed with the Modules for Experiments in Stellar Astrophysics (MESA) \citep{2011ApJS..192....3P, 2013ApJS..208....4P, 2015ApJS..220...15P, 2016ApJS..222....8D, 2016ApJ...823..102C}. See also \cite{2021AJ....161..189L} who used the code to estimate the age of main sequence stars comoving with a white dwarf, to compare with the results of their age-dating method.

Since our three white dwarfs are in \cite{2021MNRAS.508.3877G}, we had the required parameters to run the code using \cite{2018ApJ...866...21C} initial-to-final mass relation. As we don't know the composition of our white dwarfs (we are only working from the photometry), we ran the code using the derived {\it T}$_{eff}$ and log {\it g} for DA and DB white dwarfs, provided by Gentile Fusillo. The results are presented in Table \ref{tab:wd_ages}. The final mass estimates are more or less consistent with those from Gentile Fusillo. Since the white dwarf of system \#25 has a very low-mass (0.256$\pm$0.159 M$_\odot$), which is outside the range of the initial-to-final mass relation described in \cite{2018ApJ...866...21C}, \texttt{wdwarfdate} cannot estimate the main sequence age and mass of the progenitor star. However, the code can estimate the mass and cooling age of the white dwarf using \cite{2020ApJ...901...93B} cooling tracks.

\subsection{Finder Charts}

\begin{figure}[H]
\centering
\includegraphics[width=18cm]{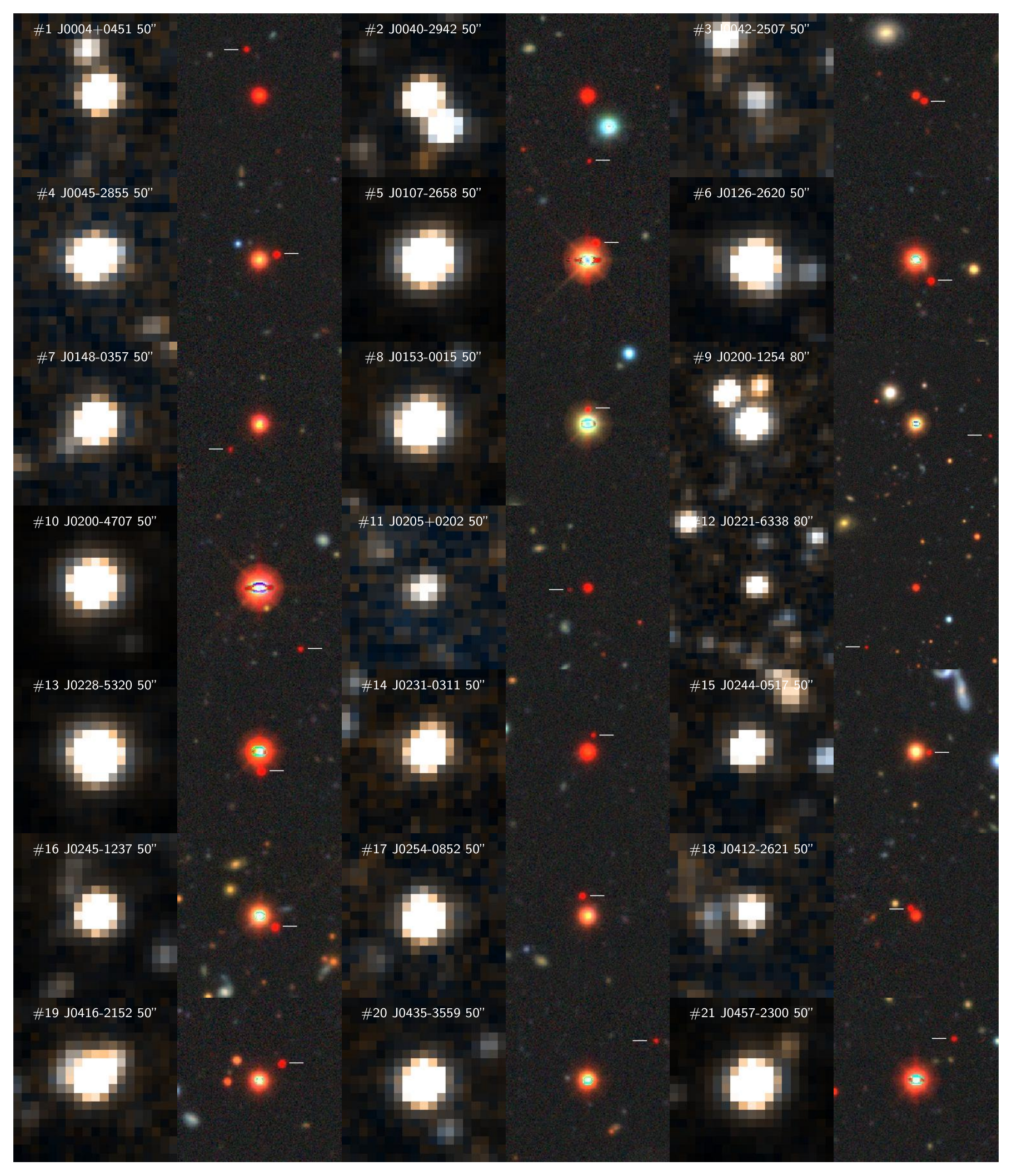}
\caption{unWISE NEO7 versus LS DR9 images (north up, east to the left). We used VISTA VHS DR6 images for systems \#27 and \#28 since there is no corresponding LS DR9 imagery. The primaries are in the center of the images, the secondaries were marked by a white dash. White lettering indicates the system number and name followed by the size of the field of view ($50\arcsec\times50\arcsec, 80\arcsec\times80\arcsec$ or $120\arcsec\times120\arcsec$, depending on the angular separation of the system components).}\label{fig:images}
\end{figure}

Figure \ref{fig:images} compares color composites from unWISE NEO7 with DESI Legacy Imaging Surveys (LS) DR9. The purpose of these images is to show the difference between low-resolution surveys such as WISE (FWHM $\sim 6 \arcsec$ for the short channels, \citealt{2010AJ....140.1868W}) compared to high-resolution surveys such as DES (FWHM $\sim 1 \arcsec$, \citealt{2018ApJS..239...18A}) or VISTA VHS (FWHM $\sim 0.9 \arcsec$, \citealt{2015A&A...575A..25S}) when it comes to identifying binary systems that are less widely separated, in which case WISE would only show one blended source rather than the resolved pair.

\onecolumngrid

\begin{figure}[H]
\centering
\includegraphics[width=18cm]{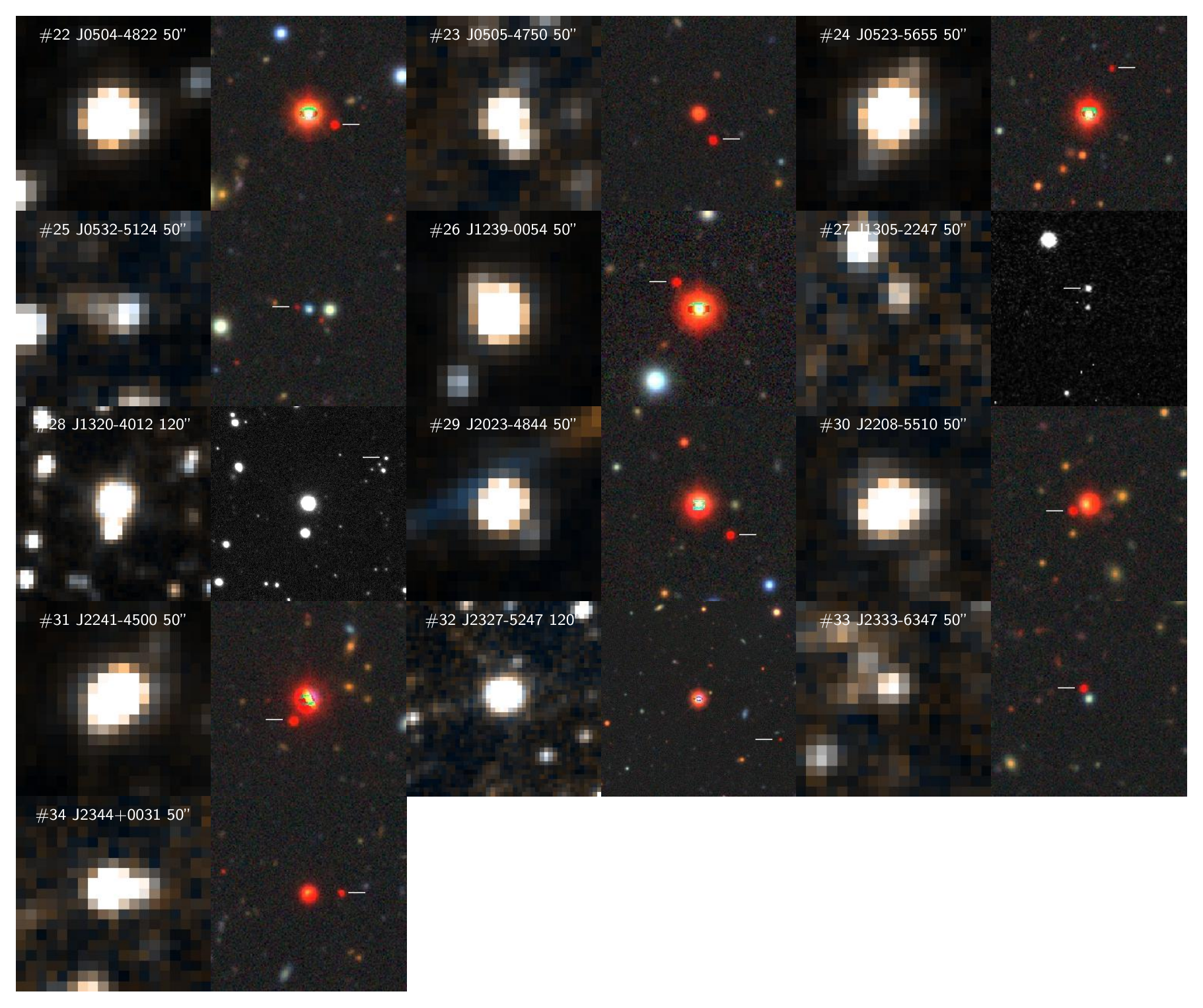}
\textbf{Figure \ref{fig:images}} (Continued)
\end{figure}

\begin{rotatepage}

\begin{longrotatetable}
\vspace*{-1.5cm}
\movetabledown=1.5cm
\begin{deluxetable*}{rllDDDDDrrrcc}
\tablecaption{Discovered systems\label{tab:systems}}
\tablehead{
\colhead{Sys.} &
\colhead{System name} &
\colhead{Short name} &
\twocolhead{$\varpi$} &
\twocolhead{$\mu_{\alpha}$} &
\twocolhead{$\mu_{\delta}$} &
\twocolhead{Dist.$^a$} &
\twocolhead{$\upsilon_{tan}$$^{b}$} &
\colhead{Sep.$^c$} &
\colhead{Sep.$^d$} &
\colhead{Pos.$^e$} &
\colhead{Sys.$^f$} &
\colhead{Ref.$^g$}
\\
\colhead{\#} &
\colhead{} &
\colhead{} &
\twocolhead{(mas)} &
\twocolhead{(mas yr$^{-1}$)} &
\twocolhead{(mas yr$^{-1}$)} &
\twocolhead{(pc)} &
\twocolhead{(km $s^{-1}$)} &
\colhead{(\arcsec)} &
\colhead{(AU)} &
\colhead{(deg)} &
\colhead{type} &
\colhead{}
}
\tabletypesize{\scriptsize}
\startdata
 & \textsc{\footnotesize Newly discovered systems} \\
\hline
1 & NSC J000456.31\ps045117.22 AB & NSC J0004\ps0451 & 9.2747$\pm$ & 0.1585 & 100.538$\pm$ & 0.206 & 10.524$\pm$ & 0.146 & 107.821$\pm$ & 1.843 & 51.663$\pm$ & 0.889 & 14.548 & 1568.531 & 15.4 & M6$\ps$L2 & \nodata \\
2 & NSC J004058.68\ms294212.37 AB & NSC J0040\ms2942 & 18.1726$\pm$ & 0.1292 & 84.313$\pm$ & 0.128 & -124.955$\pm$ & 0.143 & 55.028$\pm$ & 0.391 & 39.318$\pm$ & 0.282 & 19.906 & 1095.379 & 181.2 & M7$\ps$T5 & \nodata \\
3 & NSC J004239.82\ms250714.01 AB & NSC J0042\ms2507 & 3.8355$\pm$ & 0.8464 & 109.022$\pm$ & 0.936 & -17.709$\pm$ & 0.846 & 260.725$\pm$ & 57.535 & 136.499$\pm$ & 30.144 & 3.000 & 782.216 & 237.1 & M6$\ps$M6 & \nodata \\
4 & NSC J004509.91\ms285557.20 AB & NSC J0045\ms2855 & 6.8596$\pm$ & 0.0544 & 93.307$\pm$ & 0.055 & -24.9$\pm$ & 0.061 & 145.78$\pm$ & 1.156 & 66.732$\pm$ & 0.531 & 5.447 & 794.110 & 286.4 & M4$\ps$M9 & \nodata \\
5 & NSC J010717.67\ms265848.32 AB & NSC J0107\ms2658 & 11.5915$\pm$ & 0.0202 & 115.191$\pm$ & 0.021 & -119.037$\pm$ & 0.027 & 86.27$\pm$ & 0.15 & 67.736$\pm$ & 0.118 & 5.846 & 504.352 & 333.8 & M1$\ps$M8 & \nodata \\
6 & NSC J012629.63\ms262023.18 AB & NSC J0126\ms2620 & 6.8549$\pm$ & 0.0307 & 101.256$\pm$ & 0.04 & -51.201$\pm$ & 0.027 & 145.882$\pm$ & 0.653 & 78.459$\pm$ & 0.352 & 7.921 & 1155.583 & 215.7 & M3$\ps$M8 & \nodata \\
7 & NSC J014827.71\ms035716.63 AB & NSC J0148\ms0357 & 8.4201$\pm$ & 0.0628 & 52.933$\pm$ & 0.081 & -175.686$\pm$ & 0.042 & 118.764$\pm$ & 0.886 & 103.292$\pm$ & 0.771 & 11.752 & 1395.736 & 131.3 & M4$\ps$L4 & \nodata \\
\enddata
\hfill \\
\textsc{Note---}
The parallaxes and proper motions are from {\it Gaia} EDR3.
The primary of system \#8 and the secondary of system \#21 have confirmed spectral types of M0 and T3, respectively (see spectra in Figure \ref{fig:spectra}).
$^{(a)}$Distance, derived from the {\it Gaia} EDR3 parallax of the primary;
$^{(b)}$Tangential velocity, calculated using the {\it Gaia} EDR3 astrometry of the primary;
$^{(c)}$Angular separation;
$^{(d)}$Projected physical separation, calculated using the {\it Gaia} EDR3 parallax of the primary;
$^{(e)}$Position angle (east of north) of the secondary with respect to the primary;
$^{(f)}$System type, based on spectral type estimates (A$\ps$B);
$^{(g)}$Binary system references: (1) \cite{2020ApJS..247...66H}, (2) \cite{2019MNRAS.485.4423S}, (3) \cite{2019AA...627A.167C}
$^{(c,e)}$Calculated using NSC DR2 positions, translating the position of the primary to the epoch of the secondary.
\end{deluxetable*}
\end{longrotatetable}

\begin{longrotatetable}
\vspace*{-1cm}
\movetabledown=1cm
\begin{deluxetable*}{rlDDcDDDDDD}
\tablecaption{NSC DR2 astrometry\label{tab:nsc_astro}}
\tablehead{
\colhead{Sys.} &
\colhead{C.$^a$} &
\twocolhead{$\alpha$} &
\twocolhead{$\delta$} &
\colhead{Epoch} &
\twocolhead{$\mu_{\alpha}$} &
\twocolhead{$\mu_{\delta}$} &
\twocolhead{$\mu_{tot}$} &
\twocolhead{$\Delta\mu_{\alpha}$$^b$} &
\twocolhead{$\Delta\mu_{\delta}$$^c$} &
\twocolhead{$\Delta\mu_{tot}$$^d$}
\\
\colhead{\#} &
\colhead{} &
\twocolhead{(deg)} &
\twocolhead{(deg)} &
\colhead{} &
\twocolhead{(mas yr$^{-1}$)} &
\twocolhead{(mas yr$^{-1}$)} &
\twocolhead{(mas yr$^{-1}$)} &
\twocolhead{(mas yr$^{-1}$)} &
\twocolhead{(mas yr$^{-1}$)} &
\twocolhead{(mas yr$^{-1}$)}
}
\tabletypesize{\scriptsize}
\startdata
1 & A & 1.2346196$\pm$ & 0\farcs002078 & 4.8547833$\pm$ & 0\farcs002548 & 2017-02-17 & 101.586$\pm$ & 2.081 & 16.224$\pm$ & 2.243 & 102.873$\pm$ & 2.085 & & & & & & \\
 & B & 1.2356978$\pm$ & 0\farcs008701 & 4.8586783$\pm$ & 0\farcs008999 & 2017-01-31 & 103.442$\pm$ & 10.176 & 8.665$\pm$ & 10.271 & 103.804$\pm$ & 10.177 & 1.856$\pm$ & 10.387 & 7.559$\pm$ & 10.513 & 7.784$\pm$ & 10.506 \\
\hline
2 & A & 10.2445114$\pm$ & 0\farcs002146 & -29.703435$\pm$ & 0\farcs002093 & 2016-09-23 & 85.588$\pm$ & 1.684 & -127.509$\pm$ & 1.637 & 153.57$\pm$ & 1.652 & & & & & & \\
 & B & 10.2443636$\pm$ & 0\farcs015625 & -29.7089511$\pm$ & 0\farcs015601 & 2016-05-22 & 76.615$\pm$ & 11.596 & -137.72$\pm$ & 11.576 & 157.596$\pm$ & 11.581 & 8.973$\pm$ & 11.718 & 10.211$\pm$ & 11.691 & 13.593$\pm$ & 11.703 \\
\hline
3 & A & 10.6659165$\pm$ & 0\farcs002078 & -25.1205574$\pm$ & 0\farcs002254 & 2017-06-23 & 110.017$\pm$ & 1.081 & -18.631$\pm$ & 1.168 & 111.583$\pm$ & 1.084 & & & & & & \\
 & B & 10.6651479$\pm$ & 0\farcs002717 & -25.1210108$\pm$ & 0\farcs002906 & 2017-08-07 & 114.399$\pm$ & 1.451 & -16.276$\pm$ & 1.542 & 115.551$\pm$ & 1.453 & 4.382$\pm$ & 1.809 & 2.355$\pm$ & 1.934 & 4.975$\pm$ & 1.838 \\
\hline
4 & A & 11.2912826$\pm$ & 0\farcs002091 & -28.9325546$\pm$ & 0\farcs002165 & 2016-10-10 & 91.114$\pm$ & 1.456 & -24.391$\pm$ & 1.525 & 94.322$\pm$ & 1.461 & & & & & & \\
 & B & 11.2896207$\pm$ & 0\farcs002027 & -28.9321265$\pm$ & 0\farcs002127 & 2016-08-29 & 98.791$\pm$ & 1.421 & -25.35$\pm$ & 1.474 & 101.992$\pm$ & 1.424 & 7.677$\pm$ & 2.034 & 0.959$\pm$ & 2.121 & 7.737$\pm$ & 2.036 \\
\enddata
\hfill \\
\textsc{Note---}
$^{(a)}$Component, where A is the primary and B the secondary of the system;
$^{(b)}$Proper motion difference in right ascension;
$^{(c)}$Proper motion difference in declination;
$^{(d)}$Difference in total proper motion.
\end{deluxetable*}
\end{longrotatetable}

\begin{longrotatetable}
\vspace*{-1cm}
\movetabledown=1cm
\begin{deluxetable*}{rrrrDDrDDDDDllc}
\tablecaption{Comoving probabilities and photometric distances of the secondaries\label{tab:comov_prob}}
\tablehead{
\colhead{Sys.} &
\colhead{Comoving} &
\colhead{Comoving} &
\colhead{Comoving} &
\twocolhead{$\Delta\mu_{\alpha}$} &
\twocolhead{$\Delta\mu_{\delta}$} &
\colhead{Sep.} &
\twocolhead{Parallactic} &
\twocolhead{Mean phot.} &
\twocolhead{$\Delta_{dist.}$} &
\twocolhead{Min. phot.} &
\twocolhead{Max. phot.} &
\colhead{Band min.} &
\colhead{Band max.} &
\colhead{SpT}
\\
\colhead{\#} &
\colhead{prob. (\%)} &
\colhead{prob. (\%)} &
\colhead{prob. (\%)} &
\twocolhead{(mas yr$^{-1}$)} &
\twocolhead{(mas yr$^{-1}$)} &
\colhead{(AU)} &
\twocolhead{dist. (pc)} &
\twocolhead{dist. (pc)} &
\twocolhead{(pc)} &
\twocolhead{dist. (pc)} &
\twocolhead{dist. (pc)} &
\colhead{phot. dist.} &
\colhead{phot. dist.} &
\colhead{phot. dist.}
\\
\colhead{(1)} &
\colhead{(2)} &
\colhead{(3)} &
\colhead{(4)} &
\twocolhead{(5)} &
\twocolhead{(6)} &
\colhead{(7)} &
\twocolhead{(8)} &
\twocolhead{(9)} &
\twocolhead{(10)} &
\twocolhead{(11)} &
\twocolhead{(12)} &
\colhead{(13)} &
\colhead{(14)} &
\colhead{(15)}
}
\tabletypesize{\scriptsize}
\startdata
1 & 99.8 & 100.0 & 98.5 & 1.856$\pm$ & 10.387 & 7.559$\pm$ & 10.513 & 1569 & 107.8$\pm$ & 1.8 & 101.9$\pm$ & 8.4 & 6$\pm$ & 8.6 & 99.5$\pm$ & 11 & 104.3$\pm$ & 12.6 & $z_{NSC}$ & $i_{NSC}$ & L2 \\
2 & 99.8 & 100.0 & 100.0 & 8.973$\pm$ & 11.718 & 10.211$\pm$ & 11.691 & 1095 & 55$\pm$ & 0.4 & 42.4$\pm$ & 3.5 & 12.6$\pm$ & 3.5 & 40.8$\pm$ & 4.9 & 44.1$\pm$ & 5 & $W1_{CWISE}$ & $W2_{CWISE}$ & T5 \\
3 & 100.0 & 100.0 & 100.0 & 4.382$\pm$ & 1.809 & 2.355$\pm$ & 1.934 & 782 & 260.7$\pm$ & 57.5 & 282.5$\pm$ & 29.9 & 21.8$\pm$ & 64.8 & 263.4$\pm$ & 36.5 & 301.7$\pm$ & 47.3 & $y_{PS1}$ & $i_{PS1}$ & M6 \\
4 & 100.0 & 100.0 & 99.9 & 7.677$\pm$ & 2.034 & 0.959$\pm$ & 2.121 & 794 & 145.8$\pm$ & 1.2 & 148.9$\pm$ & 25 & 3.1$\pm$ & 25 & 148.7$\pm$ & 34.2 & 149.2$\pm$ & 36.4 & $RP_{EDR3}$ & $G_{EDR3}$ & M9 \\
5 & 99.8 & 99.9 & 99.8 & 2.888$\pm$ & 13.911 & 17.449$\pm$ & 12.082 & 504 & 86.3$\pm$ & 0.2 & 81.6$\pm$ & 12.4 & 4.7$\pm$ & 12.4 & 80$\pm$ & 17.7 & 83.2$\pm$ & 17.3 & $i_{DES}$ & $z_{DES}$ & M8 \\
6 & 99.6 & 99.8 & 99.1 & 9.646$\pm$ & 5.189 & 13.974$\pm$ & 4.68 & 1156 & 145.9$\pm$ & 0.7 & 147.2$\pm$ & 21.1 & 1.3$\pm$ & 21.1 & 145.1$\pm$ & 28.8 & 149.2$\pm$ & 30.9 & $y_{PS1}$ & $z_{PS1}$ & M8 \\
7 & 100.0 & 98.8 & 100.0 & 19.526$\pm$ & 16.492 & 2.815$\pm$ & 16.64 & 1396 & 118.8$\pm$ & 0.9 & 72.3$\pm$ & 14.4 & 46.4$\pm$ & 14.4 & 71$\pm$ & 25.4 & 73.7$\pm$ & 13.5 & $r_{DES}$ & $i_{DES}$ & L4 \\
8 & 100.0 & 100.0 & 100.0 & 0.339$\pm$ & 3.879 & 0.23$\pm$ & 4.933 & 776 & 184.5$\pm$ & 0.8 & 129.3$\pm$ & 19.6 & 55.2$\pm$ & 19.6 & 125.5$\pm$ & 27.8 & 133.1$\pm$ & 27.6 & $i_{DES}$ & $z_{DES}$ & M8 \\
\enddata
\vspace{1mm}
\begin{adjustwidth}{5mm}{}
\textbf{Column descriptions:} \\
(1) System number \\
(2) Comoving probability without a distance constraint on the secondary \\
(3) Comoving probability with a distance constraint on the secondary \\
(4) Comoving probability calculated by increasing the secondary's distance by 40\% (see Section \ref{sec:chance_align}) \\
(5) Proper motion difference in right ascension \\
(6) Proper motion difference in declination \\
(7) Projected physical separation \\
(8) Parallactic distance of the primary \\
(9) Mean photometric distance of the secondary \\
(10) Distance difference between the primary (parallactic distance) and secondary (mean photometric distance) \\
(11) Minimum photometric distance of the secondary for a given photometric system \\
(12) Maximum photometric distance of the secondary for a given photometric system \\
(13) Band from which the minimum photometric distance of the secondary was inferred (CWISE and UWISE stand for CatWISE2020 and unWISE DR1, respectively) \\
(14) Band from which the maximum photometric distance of the secondary was inferred \\
(15) Spectral type estimate used to derive the absolute magnitude for the photometric distance calculations
\end{adjustwidth}
\end{deluxetable*}
\end{longrotatetable}

\end{rotatepage}

\onecolumngrid

\begin{deluxetable}{rlDDDDDcc}[H]
\tablecaption{Fundamental parameters and literature references\label{tab:fund_param}}
\tablehead{
\colhead{Sys.} &
\colhead{Component} &
\twocolhead{{\it T}$_{eff}$} &
\twocolhead{log {\it g}} &
\twocolhead{Mass} &
\twocolhead{Radius} &
\twocolhead{Luminosity} &
\colhead{Param.} &
\colhead{Other}
\\
\colhead{\#} &
\colhead{name} &
\twocolhead{(K)} &
\twocolhead{} &
\twocolhead{(M$_\odot$)} &
\twocolhead{(R$_\odot$)} &
\twocolhead{(L$_\odot$)} &
\colhead{ref.$^c$} &
\colhead{ref.$^d$}
}
\tabletypesize{\footnotesize}
\startdata
1 & NSC J0004\ps0451 A & 2900$\pm$ & 157 & 5.0627$\pm$ & 0.0207 & 0.17$\pm$ & 0.021 & 0.201$\pm$ & 0.007 & 0.00257$\pm$ & 0.00075 & 1 & \\
1 & NSC J0004\ps0451 B & \nodata & & \nodata & & \nodata & & \nodata & & \nodata & & & 3 \\
2 & NSC J0040\ms2942 A & 2823$\pm$ & 157 & 5.2584$\pm$ & 0.0648 & 0.097$\pm$ & 0.02 & 0.121$\pm$ & 0.004 & 0.00084$\pm$ & 0.00024 & 1 & 4,5 \\
4 & NSC J0045\ms2855 A & 3326$\pm$ & 157 & 4.9128$\pm$ & 0.0029 & 0.29$\pm$ & 0.021 & 0.312$\pm$ & 0.01 & 0.01073$\pm$ & 0.00273 & 1 & \\
5 & NSC J0107\ms2658 A & 3677$\pm$ & 157 & 4.7644$\pm$ & 0.007 & 0.467$\pm$ & 0.02 & 0.469$\pm$ & 0.014 & 0.03627$\pm$ & 0.0084 & 1 & 5,6,7,8 \\
6 & NSC J0126\ms2620 A & 3435$\pm$ & 157 & 4.7987$\pm$ & 0.0058 & 0.426$\pm$ & 0.021 & 0.431$\pm$ & 0.013 & 0.02327$\pm$ & 0.00573 & 1 & \\
7 & NSC J0148\ms0357 A & 3343$\pm$ & 158 & 4.9799$\pm$ & 0.0109 & 0.226$\pm$ & 0.02 & 0.255$\pm$ & 0.008 & 0.00731$\pm$ & 0.00187 & 1 & 6,9 \\
8 & NSC J0153\ms0015 A & 3962$\pm$ & 157 & 4.6746$\pm$ & 0.0102 & 0.565$\pm$ & 0.02 & 0.572$\pm$ & 0.017 & 0.07275$\pm$ & 0.01593 & 1 & \\
\enddata
\hfill \\
$^{(a)}${\it T}$_{eff}$, log {\it g} and mass for DA white dwarfs \\
$^{(b)}${\it T}$_{eff}$, log {\it g} and mass for DB white dwarfs \\
$^{(c)}$Parameter references: (1) \cite{2019AJ....158..138S}, (2) \cite{2021MNRAS.508.3877G} \\
$^{(d)}$Other literature references: (3) \cite{2016AA...589A..49S}, (4) \cite{2018AA...619L...8R}, (5) \cite{2003AA...397..575P}, (6) \cite{1995yCat.1098....0L}, (7) \cite{1976ApJS...30..351E}, (8) \cite{1976ApJ...204..101E}, (9) \cite{2003ApJ...582.1011S}, (10) \cite{1996AAS..115..481W}, (11) \cite{2016ApJS..224...36K}
\end{deluxetable}

\begin{deluxetable}{rlrrrrrrc}[H]
\tablecaption{Total masses, mass ratios and binding energies\label{tab:mass_ratios}}
\tablehead{
\colhead{Sys.} &
\colhead{Sys.} &
\colhead{Sep.$^a$} &
\colhead{M$_{1}$$^b$} &
\colhead{M$_{2}$$^c$} &
\colhead{M$_{tot}$$^d$} &
\colhead{M$_{2}$/M$_{1}$$^e$} &
\colhead{E$_{bin}$$^f$} &
\colhead{SpT$_{2}$$^g$}
\\
\colhead{\#} &
\colhead{name} &
\colhead{(AU)} &
\colhead{(M$_\odot$)} &
\colhead{(M$_\odot$)} &
\colhead{(M$_\odot$)} &
\colhead{} &
\colhead{($\times 10^{41}$ erg)} &
\colhead{}
}
\tabletypesize{\footnotesize}
\startdata
1 & NSC J0004$\ps$0451 & 1568.531 & 0.170 & 0.073 & 0.243 & 0.427 & 1.11 & L2 \\
2 & NSC J0040$\ms$2942 & 1095.379 & 0.097 & 0.034 & 0.131 & 0.354 & 0.42 & T5 \\
3 & NSC J0042$\ms$2507 & 782.216 & 0.128 & 0.115 & 0.243 & 0.901 & 2.64 & M6 \\
4 & NSC J0045$\ms$2855 & 794.110 & 0.290 & 0.079 & 0.369 & 0.273 & 4.04 & M9 \\
5 & NSC J0107$\ms$2658 & 504.352 & 0.467 & 0.090 & 0.557 & 0.192 & 11.67 & M8 \\
6 & NSC J0126$\ms$2620 & 1155.583 & 0.426 & 0.090 & 0.516 & 0.211 & 4.65 & M8 \\
7 & NSC J0148$\ms$0357 & 1395.736 & 0.226 & 0.055 & 0.281 & 0.245 & 1.25 & L4 \\
8 & NSC J0153$\ms$0015 & 776.150 & 0.565 & 0.090 & 0.655 & 0.159 & 9.18 & M8 \\
\enddata
\hfill \\
\textsc{Note---}
$^{(a)}$Projected physical separation;
$^{(b)}$Primary mass estimate from either \cite{2019AJ....158..138S} or \cite{2021MNRAS.508.3877G};
$^{(c)}$Secondary mass estimate from either \cite{Mamajek2021} for sources with M$_G$ $<$ 17 mag, or from the SpT to dynamical mass values computed in \cite{Dupuy2017};
$^{(d)}$Total mass (primary mass $\ps$ secondary mass);
$^{(e)}$Mass ratio (secondary/primary);
$^{(f)}$Binding energy, calculated by multiplying the physical separation by 1.26 to account for inclination angle and eccentricity of the binary orbits \citep{1992ApJ...396..178F};
$^{(g)}$Spectral type used to derive the secondary's mass.
\end{deluxetable}

\begin{deluxetable}{rlDDDDDDDD}[H]
\tablecaption{Age, initial and final mass estimates of the white dwarfs\label{tab:wd_ages}}
\tablehead{
\colhead{Sys.} &
\colhead{WD} &
\twocolhead{{\it T}$_{eff}$$^a$} &
\twocolhead{log {\it g}$^b$} &
\twocolhead{Mass$^c$} &
\twocolhead{MS age$^d$} &
\twocolhead{WD age$^e$} &
\twocolhead{Tot. age$^f$} &
\twocolhead{Init. mass$^g$} &
\twocolhead{Final mass$^h$}
\\
\colhead{\#} &
\colhead{type} &
\twocolhead{(K)} &
\twocolhead{} &
\twocolhead{(M$_\odot$)} &
\twocolhead{(Gyr)} &
\twocolhead{(Gyr)} &
\twocolhead{(Gyr)} &
\twocolhead{(M$_\odot$)} &
\twocolhead{(M$_\odot$)}
\\
\colhead{(1)} &
\colhead{(2)} &
\twocolhead{(3)} &
\twocolhead{(4)} &
\twocolhead{(5)} &
\twocolhead{(6)} &
\twocolhead{(7)} &
\twocolhead{(8)} &
\twocolhead{(9)} &
\twocolhead{(10)}
}
\tabletypesize{\footnotesize}
\startdata
25 & DA & 4960$\pm$ & 422 & 7.225$\pm$ & 0.507 & 0.256$\pm$ & 0.159 & \nodata & & 2.34 & $_{-0.91}^{+2.04}$ & \nodata & & \nodata & & 0.33 & $_{-0.11}^{+0.21}$ \\
25 & DB & 5043$\pm$ & 337 & 7.321$\pm$ & 0.393 & 0.283$\pm$ & 0.133 & \nodata & & 2.04 & $_{-0.65}^{+1.62}$ & \nodata & & \nodata & & 0.32 & $_{-0.1}^{+0.16}$ \\
27 & DA & 5373$\pm$ & 786 & 8.076$\pm$ & 0.670 & 0.63$\pm$ & 0.383 & 1.42 & $_{-1.01}^{+2.15}$ & 5.25 & $_{-2.07}^{+2.27}$ & 7.41 & $_{-2.06}^{+1.75}$ & 1.93 & $_{-0.53}^{+1.18}$ & 0.65 & $_{-0.05}^{+0.12}$ \\
27 & DB & 5266$\pm$ & 720 & 8.003$\pm$ & 0.632 & 0.571$\pm$ & 0.355 & 1.46 & $_{-1.0}^{+2.14}$ & 5.39 & $_{-2.05}^{+2.22}$ & 7.57 & $_{-2.01}^{+1.65}$ & 1.91 & $_{-0.51}^{+1.09}$ & 0.65 & $_{-0.05}^{+0.1}$ \\
33 & DA & 5243$\pm$ & 318 & 8.067$\pm$ & 0.282 & 0.623$\pm$ & 0.175 & 1.65 & $_{-0.98}^{+2.05}$ & 5.52 & $_{-1.62}^{+1.84}$ & 7.79 & $_{-1.59}^{+1.4}$ & 1.82 & $_{-0.42}^{+0.83}$ & 0.64 & $_{-0.05}^{+0.07}$ \\
33 & DB & 5147$\pm$ & 285 & 8.002$\pm$ & 0.253 & 0.57$\pm$ & 0.143 & 1.73 & $_{-0.94}^{+1.92}$ & 5.73 & $_{-1.58}^{+1.7}$ & 8.05 & $_{-1.5}^{+1.25}$ & 1.78 & $_{-0.38}^{+0.71}$ & 0.63 & $_{-0.04}^{+0.06}$ \\
\enddata
\hfill \\
\end{deluxetable}

\vspace{-1cm}

\twocolumngrid

\section{Discussion}\label{sec:discussion}
Systems \#9 and \#19 have high tangential velocities of $\sim187$ and $\sim158$ km s$^{-1}$ respectively, suggesting a possible thick disk or halo membership \citep{2010ApJ...712..692C, 2012ApJS..201...19D, 2018MNRAS.478..611B, 2020MNRAS.492.3816A}. Further study of these systems is required to determine how old or metal-poor these objects really are.

Systems \#13, \#14 and \#31 have a primary with a high {\it Gaia} RUWE\footnote{Re-normalized Unit Weight Error} of $\sim1.9$, $\sim14$ and $\sim2.7$ respectively. A RUWE significantly greater than 1.0 (e.g. $>$1.4) could indicate that the source is non-single and thus might be an unresolved binary \citep{2020MNRAS.496.1922B, 2020MNRAS.495..321P}. Given past work suggesting that widely separated ultracool dwarfs from higher mass companions are likely to be found in hierarchical systems, there is an increased probability of finding more hidden companions. (e.g. \citealt{Burgasser2003}, \citealt{Faherty2010}, \citealt{Law2010}). System \#31, which is likely a triple system (M\ps M\ps L), has already been analyzed in more detail \citep{2021RNAAS...5..196K}.

Systems \#9, \#13 and \#15 have comoving probabilities well below 100\%, when we use a distance constraint in these calculations, indicating that they are likely chance alignments. Systems \#9 and \#15 also show the largest discrepancies between the primary's parallactic and the secondary's photometric distance (232.5$\pm$1.1 pc vs. 121.8$\pm$30.2 pc and 258$\pm$3.4 pc vs. 128.4$\pm$10.7 pc, respectively), implying that these systems (provided the photometric distance is reasonably correct) cannot be physically bound. Whereas system \#13 has a much smaller distance difference of only 8.8$\pm$14 pc between the primary and the secondary as well as a higher comoving probability (87.0\%) compared to systems \#9 and \#15 (59.8 and 4.5\%, respectively), but still well below that of the majority of the systems ($>$ 99\%).

\section{Conclusion}

We presented the discovery of 34 new, potential comoving systems found by the means of NSC DR2 and demonstrated that the catalog’s proper motions are well suited for the discovery of comoving systems including late-type objects such as brown dwarfs in the L and T regimes. We also showed that the catalog can be used to find close late-type comoving companions to white dwarfs, providing valuable benchmark systems for the difficult task of estimating brown dwarf ages.

According to the literature, none of the 34 identified objects were previously known to have a comoving companion. Some of the systems stand out as having high tangential velocities, suggesting that they could be candidates for low metallicity benchmark systems. The newly discovered systems expand the known sample of benchmark systems involving L and T dwarfs and thus contribute to the further characterization of cool substellar objects.

Spectroscopic follow-up on individual interesting systems is warranted, for instance, to find out how low the metallicity of the high tangential velocity systems actually is, or to clarify the potential binarity of the primaries with a high {\it Gaia} RUWE value.

\section{Acknowledgements}

This work has made use of data and/or services provided by:
\begin{itemize}
    \item The Astro Data Lab at NSF's National Optical-Infrared Astronomy Research Laboratory. NOIRLab is operated by the Association of Universities for Research in Astronomy (AURA), Inc. under a cooperative agreement with the National Science Foundation.
    \item The DESI Legacy Imaging Surveys (\url{https://www.legacysurvey.org/acknowledgment})
    \item The VISTA Science Archive (VSA) holding the image and catalogue data products generated by VIRCAM on the Visible and Infrared Survey Telescope for Astronomy (VISTA)
    \item The European Space Agency (ESA) mission {\it Gaia} (\url{https://www.cosmos.esa.int/gaia}), processed by the {\it Gaia} Data Processing and Analysis Consortium (DPAC, \url{https://www.cosmos.esa.int/web/gaia/dpac/consortium}). Funding for the DPAC has been provided by national institutions, in particular the institutions participating in the {\it Gaia} Multilateral Agreement.
    \item The European Southern Observatory under ESO program VHS 179.A-2010
    \item The SIMBAD database \citep{2000A&AS..143....9W}, operated at CDS, Strasbourg, France
    \item The VizieR catalogue access tool, CDS, Strasbourg, France (DOI : 10.26093/cds/vizier). The original description of the VizieR service was published in \cite{2000A&AS..143...23O}.
    \item The Infrared Telescope Facility, which is operated by the University of Hawaii under contract 80HQTR19D0030 with the National Aeronautics and Space Administration
    \item The SpeX Prism Spectral Libraries, maintained by Adam Burgasser at  \url{http://pono.ucsd.edu/~adam/browndwarfs/spexprism}
    \item The UltracoolSheet, maintained by Will Best, Trent Dupuy, Michael Liu, Rob Siverd, and Zhoujian Zhang, and developed from compilations by \cite{2012ApJS..201...19D}, \cite{2013Sci...341.1492D}, \cite{2016ApJ...833...96L}, \cite{2018ApJS..234....1B}, and \cite{2021AJ....161...42B}.
    \item \texttt{wdwarfdate}, an open source code which estimates ages of white dwarfs in a Bayesian framework from effective temperature and surface gravity \url{https://github.com/rkiman/wdwarfdate}
    \item \texttt{AstroToolBox}, a Java tool set for the identification and classification of astronomical objects with a focus on very low-mass and ultra-cool dwarfs \url{https://github.com/fkiwy/AstroToolBox}
    \newline
    \item This research was supported by National Science Foundation Grant No.’s 2007068, 2009136, and 2009177 and J.F. acknowledge support from the Heising-Simons Foundation.
\end{itemize}

\bibliography{manuscript}
\bibliographystyle{aasjournal}

\appendix
\begin{rotatepage}

\begin{longrotatetable}
\vspace*{-1cm}
\movetabledown=1cm
\begin{deluxetable*}{rlDDDDDDDDDDD}
\tablecaption{Optical and infrared photometry\label{tab:photometry}}
\tablehead{
\colhead{Sys.} &
\colhead{C.$^a$} &
\twocolhead{G$_{BP}$} &
\twocolhead{G} &
\twocolhead{G$_{RP}$} &
\twocolhead{r$_{PS1}$} &
\twocolhead{i$_{PS1}$} &
\twocolhead{z$_{PS1}$} &
\twocolhead{y$_{PS1}$} &
\twocolhead{r$_{NSC}$} &
\twocolhead{i$_{NSC}$} &
\twocolhead{z$_{NSC}$} &
\twocolhead{Y$_{NSC}$}
\\
\colhead{\#} &
\colhead{} &
\twocolhead{J$_{2M}$} &
\twocolhead{H$_{2M}$} &
\twocolhead{Ks$_{2M}$} &
\twocolhead{J$_{VHS}$} &
\twocolhead{H$_{VHS}$} &
\twocolhead{Ks$_{VHS}$} &
\twocolhead{W1$_{A}$} &
\twocolhead{W2$_{A}$} &
\twocolhead{W3$_{A}$} &
\twocolhead{W1$_{C}$} &
\twocolhead{W2$_{C}$}
\\
\colhead{} &
\colhead{} &
\twocolhead{(mag)} &
\twocolhead{(mag)} &
\twocolhead{(mag)} &
\twocolhead{(mag)} &
\twocolhead{(mag)} &
\twocolhead{(mag)} &
\twocolhead{(mag)} &
\twocolhead{(mag)} &
\twocolhead{(mag)} &
\twocolhead{(mag)} &
\twocolhead{(mag)}
}
\tabletypesize{\scriptsize}
\startdata
1 & A & 20.314$\pm$ & 0.065 & 17.904$\pm$ & 0.003 & 16.491$\pm$ & 0.005 & 19.221$\pm$ & 0.008 & 17.174$\pm$ & 0.004 & 16.203$\pm$ & 0.003 & 15.702$\pm$ & 0.004 & 19.031$\pm$ & 0.003 & 16.947$\pm$ & 0.002 & 16.005$\pm$ & 0.001 & 15.643$\pm$ & 0.002 \\
 & & 14.301$\pm$ & 0.029 & 13.688$\pm$ & 0.04 & 13.381$\pm$ & 0.034 & \nodata & & \nodata & & \nodata & & 13.208$\pm$ & 0.025 & 12.995$\pm$ & 0.03 & $>$12.226 & & 13.2$\pm$ & 0.013 & 12.995$\pm$ & 0.011 \\
 & B & \nodata & & \nodata & & \nodata & & 22.159$\pm$ & 0.343 & 21.797$\pm$ & 0.183 & 20.609$\pm$ & 0.129 & 19.641$\pm$ & 0.052 & \nodata & & 21.821$\pm$ & 0.033 & 20.288$\pm$ & 0.015 & 19.696$\pm$ & 0.03 \\
 & & \nodata & & \nodata & & \nodata & & \nodata & & \nodata & & \nodata & & 15.278$\pm$ & 0.039 & 15.137$\pm$ & 0.1 & $>$12.551 & & 15.373$\pm$ & 0.021 & 15.138$\pm$ & 0.031 \\
\hline
2 & A & 20.79$\pm$ & 0.103 & 17.983$\pm$ & 0.003 & 16.495$\pm$ & 0.006 & 19.822$\pm$ & 0.015 & 17.256$\pm$ & 0.005 & 16.147$\pm$ & 0.003 & 15.566$\pm$ & 0.004 & 19.519$\pm$ & 0.004 & 16.944$\pm$ & 0.002 & 15.879$\pm$ & 0.001 & 15.528$\pm$ & 0.002 \\
 & & 14.118$\pm$ & 0.03 & 13.576$\pm$ & 0.029 & 13.307$\pm$ & 0.033 & \nodata & & \nodata & & \nodata & & 13.004$\pm$ & 0.025 & 12.788$\pm$ & 0.026 & $>$12.514 & & \nodata & & \nodata & \\
 & B & \nodata & & \nodata & & \nodata & & \nodata & & \nodata & & \nodata & & \nodata & & \nodata & & \nodata & & 21.238$\pm$ & 0.033 & 20.145$\pm$ & 0.044 \\
 & & \nodata & & \nodata & & \nodata & & \nodata & & \nodata & & \nodata & & \nodata & & \nodata & & \nodata & & 17.575$\pm$ & 0.075 & 15.91$\pm$ & 0.057 \\
\enddata
\hfill \\
\textsc{Note---}
$^{(a)}$Component, where A is the primary and B the secondary of the system. \\
{\it Gaia} photometry is from EDR3, Pan-STARRS and NSC photometry is from DR2, VHS photometry is from DR5. \\
The photometric band subscripts \emph{2M}, \emph{A} and \emph{C} stand for 2MASS, AllWISE and CatWISE, respectively.
\end{deluxetable*}
\end{longrotatetable}

\end{rotatepage}

\onecolumngrid

\startlongtable
\begin{deluxetable}{lc|r>{\bfseries}rrrr|r>{\bfseries}rrrr}
\tablecaption{Mean photometric uncertainties per band and color\label{tab:phot_errors}}
\tablehead{
\multicolumn{2}{c}{} \vline &
\multicolumn{5}{c}{\textbf{Primary photometric uncertainties}} \vline &
\multicolumn{5}{c}{\textbf{Secondary photometric uncertainties}}
\\
\cline{3-7}
\cline{8-12}
\colhead{Catalog} &
\colhead{Band} \vline &
\colhead{Number} &
\colhead{\textbf{Mean}} &
\colhead{Min} &
\colhead{Median} &
\colhead{Max} \vline &
\colhead{Number} &
\colhead{\textbf{Mean}} &
\colhead{Min} &
\colhead{Median} &
\colhead{Max}
\\
\cline{2-2}
\colhead{} &
\colhead{Color} \vline &
\colhead{of objects} &
\colhead{\textbf{(mag)}} &
\colhead{(mag)} &
\colhead{(mag)} &
\colhead{(mag)} \vline &
\colhead{of objects} &
\colhead{\textbf{(mag)}} &
\colhead{(mag)} &
\colhead{(mag)} &
\colhead{(mag)}
}
\tabletypesize{\footnotesize}
\startdata
Gaia EDR3 & BP & 34 & 0.046 & 0.003 & 0.009 & 0.244 & \nodata & \nodata & \nodata & \nodata & \\
Gaia EDR3 & G & 34 & 0.004 & 0.003 & 0.003 & 0.008 & 5 & 0.017 & 0.010 & 0.018 & 0.023 \\
Gaia EDR3 & RP & 34 & 0.014 & 0.004 & 0.004 & 0.115 & 5 & 0.100 & 0.048 & 0.089 & 0.167 \\
SDSS DR16 & g & 10 & 0.050 & 0.003 & 0.017 & 0.378 & \nodata & \nodata & \nodata & \nodata & \\
SDSS DR16 & r & 10 & 0.019 & 0.003 & 0.010 & 0.119 & \nodata & \nodata & \nodata & \nodata & \\
SDSS DR16 & i & 10 & 0.006 & 0.001 & 0.005 & 0.021 & 5 & 0.156 & 0.051 & 0.150 & 0.330 \\
SDSS DR16 & z & 10 & 0.007 & 0.003 & 0.005 & 0.024 & 5 & 0.114 & 0.064 & 0.101 & 0.198 \\
PS1 DR2 & g & 18 & 0.012 & 0.001 & 0.007 & 0.047 & \nodata & \nodata & \nodata & \nodata & \\
\enddata
\hfill \\
\textsc{Note---}
The mean uncertainties of the colors correspond to the propagated mean uncertainties in the respective bands. 
\end{deluxetable}

\end{document}